\documentclass{article}
\usepackage{spconf}
\usepackage[utf8]{inputenc}	%
\usepackage{url} %
\usepackage{siunitx} %
\usepackage{float} %
\usepackage[pdftex]{graphicx} %

\usepackage{amsmath}  %
\usepackage{amssymb} %

\newcommand{\dissim}[0]{d} %
\newcommand{\crossvalk}[0]{5} %

\newcommand{\centi}[0]{j} %
\newcommand{\filti}[0]{i} %
\newcommand{\optpatch}[0]{\mathbf{o}} %
\newcommand{\iband}[0]{b} %
\newcommand{\nband}[0]{B} %
\newcommand{\ipix}[0]{k} %
\newcommand{\npix}[0]{N} %
\newcommand{\pixpatch}[0]{\mathbb{P}} %

\title{Polarimetric Guided Nonlocal Means Covariance Matrix Estimation for Defoliation Mapping}

\name{Jørgen A.\ Agersborg$^{\dagger}$, Stian Normann Anfinsen$^{\dagger}$ and Jane Uhd Jepsen$^{\ddagger}$}
\address{$^{\dagger}$UiT The Arctic University of Norway, Department of Physics and Technology, Tromsø, Norway \\
$^{\ddagger}$ Norwegian Institute for Nature Research, Tromsø, Norway}

\begin{document}

\maketitle

\section{Abstract}
\label{sec:abstract}
In this study we investigate the potential for using synthetic aperture radar (SAR) data to provide high resolution defoliation and regrowth mapping of trees in the tundra-forest ecotone. Using aerial photographs, four areas with live forest and four areas with dead trees were identified. Quad-polarimetric SAR data from RADARSAT-2 was collected from the same area, and the complex multilook polarimetric covariance matrix was calculated using a novel extension of guided nonlocal means speckle filtering. The nonlocal approach allows us to preserve the high spatial resolution of single-look complex data, which is essential for accurate mapping of the sparsely scattered trees in the study area. Using a standard random forest classification algorithm, our filtering results in over $99.7 \%$ classification accuracy, higher than traditional speckle filtering methods, and on par with the classification accuracy based on optical data.

\section{Introduction}
\label{sec:intro}

The tundra-forest ecotone is the boundary between the low arctic tundra and the subarctic forest.
A warming climate is expected to lead to encroachment of woody species into the tundra, however this will be counteracted locally by herbivores such as browsing ungulates or defoliating forest pest insects.
Geometrid moth outbreaks cause defoliation and tree mortality, and can lead to rapid state transitions of the tundra-forest ecotone.
Defoliating species such as geometrids usually do not kill their host tree outright, but inflict damage that accumulated over several years, and often in combination with other stressors, leads to an increase in tree mortality \cite{jepsen2013ecosystem, senf2017remote}.
The outbreaks affect large areas, but recovery of the crown layer of the birch forest is highly dependent on local factors such as ungulate browsing, soil moisture and quality.

Remote sensing from satellites provides a valuable contribution by being able to monitor the effects of birch moth outbreaks and the regrowth after for vast areas. The approach taken in previous work is to detect defoliation based on coarse resolution (pixel resolution $> 200 m$) normalised difference vegetation index (NDVI) products derived from multispectral optical remote sensing images, and correlating this with field work measurements of larvae densities \cite{jepsen2009monitoring}. This follows the convention that defoliation studies often are based on NDVI products. A literature review published in 2017 shows that 82 \% of studies mapping defoliation of broadleaved forest caused by insect disturbance used a single spectral index, and most frequently NDVI \cite{senf2017remote}.

In this work, we will consider synthetic aperture radar (SAR) for primarily three reasons. Firstly, polarimetric SAR data are theoretically able to differentiate between scattering mechanisms such as surface, volume, and double bounce. Hence it could be able to accurately separate live tree crowns (volume scattering) from defoliated trees (double bounce scattering). Secondly, remote sensing products from satellite based SAR are near weather-independent. The Norwegian low arctic tundra has a high average cloud cover percentage, which limits observations by optical satellites. And thirdly, it would be interesting to evaluate how SAR performs when it comes to monitor defoliation. While SAR has been used to monitor \emph{deforestation}, none of the studies of broadleaved forest \emph{defoliation} summarised in \cite{senf2017remote} used SAR.

For remote sensing to contribute to understanding the complicated dynamics of the tundra-forest ecotone, it is important that it manages to separate between areas with live and defoliated crown in a setting where trees are sparse and these two classes are interwoven on a fine scale. This leads to the stringent requirement that we would like to preserve as much of the spatial resolution as possible. This again inspired us to extend the guided nonlocal means (GNLM) speckle filtering algorithm \cite{vitale2019gnlm} to estimate complex covariance matrices, preserving the spatial resolution of single-look complex data. A random forest classifier was then employed on the filtered covariance matrices to separate live from defoliated pixels.

\section{Data collection and preprocessing}
\label{sec:method}

\begin{figure}[t]
\begin{minipage}[b]{1.0\linewidth}
  \centering
  \includegraphics[height=8.0cm, width=1.0\linewidth]{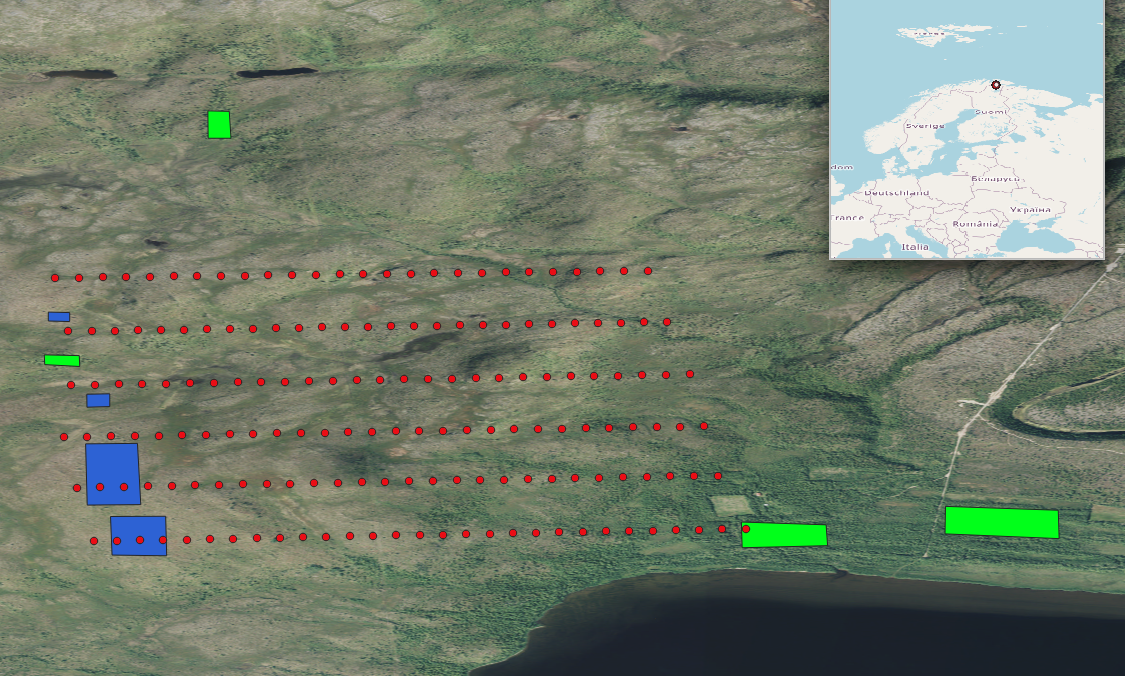}
\end{minipage}
\caption{Overview of the Polmak study area with reference areas (rectangles) and transects (red circles).}
\label{fig:studysite}
\end{figure}

The study area is close to Polmak, Norway and Nuorgam, Finland in an area of the subarctic birch forest which stretches across the Norwegian-Finnish border. The effects on the forest of a major birch moth outbreak between 2006 and 2008 are still clearly visible.
By studying high resolution aerial photographs, and comparing images from before (2005) and after the outbreak (2010), eight reference areas (RAs) were identified%
. All RAs were forested before the outbreak, but four of the eight areas had no live canopy after the outbreak. These were classified as dead and defoliated forest, marked as blue rectangles in Fig.\ \ref{fig:studysite}. The remaining four RAs, marked in green, represent the live forest class.
During fieldwork in 2017, detailed measurements were done for 165 $10\mathrm{m}\times10\mathrm{m}$ ground plots (red dots in Fig.\ \ref{fig:studysite}). Six of these are inside three of the RAs. These measurements, while few and not systematically sampled with respect to the RAs, indicate that the classes based on the aerial photographs are correctly set.

Two fine resolution quad-polarisation RADARSAT-2 scenes from July 25th and August 1st 2017 were obtained. The nominal scene size is $25\mathrm{km}\times25\mathrm{km}$ with a nominal resolution of $5.2\mathrm{m}\times7.6\mathrm{m}$ (range $\times$ azimuth). Each product was radiometrically calibrated and terrain corrected using the European Space Agency (ESA) Sentinel Application Platform (SNAP) software. The terrain corrected output products had $10.0\mathrm{m}\times10.0\mathrm{m}$ spatial resolution.

Next, the data were filtered to suppress the noise-like speckle phenomenon inherent to all SAR data.
A recent development in speckle filtering is the GNLM algorithm, which uses a co-registered optical image to guide the nonlocal filtering procedure \cite{vitale2019gnlm}. For the optical guide image, a Sentinel-2 image from July 26th 2017 covering the study area was obtained from the Copernicus data hub. Atmospheric correction was applied to retrieve the top of atmosphere reflectance (TOA). Then all spectral bands with $10.0\mathrm{m}$ spatial resolution, namely red, green, blue, and near infrared (NIR), were extracted. For comparison of classification performance, the NDVI was also calculated, $\text{NDVI} = (\text{NIR-red})/(\text{NIR+red})$.

Since we are interested in the different scattering phenomena, we consider the multilook complex covariance matrix, $\mathbf{C}$. For each pixel, the complex scattering vector is
\begin{equation}
  \label{eq:s-vec}
  \mathbf{s} = \left[ S_{HH}, S_{HV}, S_{VV} \right]^T \in \mathbb{C}^{3 \times 1}\,,
\end{equation}
where the subscripts indicate the polarisations of the transmitted pulse and the received polarisation (horizontal (H) or vertical (V)). We have assumed reciprocity, $S_{HV} = S_{VH}$. Given a set of scattering vectors $\mathbf{s}$, the sample covariance matrix $\mathbf{C}$ can be computed as the sample mean, $\mathbf{C} = \langle \mathbf{s} \mathbf{s}^H \rangle$,
where the brackets denote averaging and the superscript $H$ the conjugate transpose operation. In the SNAP software, the speckle filtering step is combined with the estimation of $\mathbf{C}$, where the different speckle filtering algorithms determine how the complex scattering vectors used in the average are selected.%

\section{Polarimetric Guided Nonlocal Means}
\label{sec:pgnlm}

Nonlocal algorithms are based on splitting the denoising problem into two steps; 1) Finding good predictors and 2) using these predictors in the estimation \cite{deledalle2014exploiting}.
While the boxcar algorithm makes the implicit assumption that the closest pixels make the best estimators, nonlocal algorithms uses a similarity criterion to find estimators.

Many different nonlocal algorithms have been used for speckle filtering, and often the similarity criteria are based on patch-wise similarity measures for robustness \cite{deledalle2014exploiting}. A further extension of nonlocal filtering was proposed in \cite{he2012guided}, where a guide image was used to help select similar pixels for averaging. The guidance image could be the noisy input itself, a pre-filtered version of it, or another, coregistered, image \cite{he2012guided}. An example of the former is \cite{ma2018nonlinear} which used a guided filtering framework to filter the polarimetric covariance matrices, where the guide used was the noisy SAR image itself.

The use of a coregistered optical image to guide the SAR despeckling was first proposed in \cite{verdoliva2014optical-driven-nonlocal}.
An important aspect of the methodology was that the filtered output was the combination of SAR pixels only, to avoid injecting optical image geometry into the SAR scene \cite{verdoliva2014optical-driven-nonlocal}.

Gaetano \emph{et al.\ }\cite{gaetano2017fuse-sar-opt} extended the previous work in \cite{verdoliva2014optical-driven-nonlocal} to use a nonlocal means framework, except for strongly heterogeneous areas of the SAR scene.
Also \cite{gaetano2017fuse-sar-opt} improved the earlier results by using patch-based filtering. The need to explicitly test for heterogeneous areas was replaced by reliability tests which removes unreliable predictors in a further development \cite{vitale2019gnlm}.

The previously mentioned work only deals with single-channel intensity SAR images \cite{verdoliva2014optical-driven-nonlocal, gaetano2017fuse-sar-opt, vitale2019gnlm}, where the filtering problem can be formulated as estimating the "clean" intensity image $\mathbf{\hat{X}}$ based on the original noisy intensity data $\mathbf{X}$ aided by the coregistered optical guide image $\mathbf{O}$. The filtering is patch-based, where a patch centred on a pixel with spatial index (pixel position) $\centi$ is defined as $\mathbf{x}(\centi) = \{ \mathbf{X}(\centi + \ipix) \, , \, \ipix \in \pixpatch \}$, where $\pixpatch$ indicates a set of $\npix$ spatial offsets with respect to $\centi$ \cite{vitale2019gnlm}.

The filtering is then done for each patch $\mathbf{x}(\centi)$ centred on pixel $\centi$ in the input SAR image, by summing the weighted patches $\mathbf{x}(\filti)$ in a search area $\Omega(\centi)$ around $\centi$:
\begin{equation}
  \label{eq:gnlm-pix}
  \mathbf{\hat{x}}(\centi) = \sum\limits_{\filti \in \Omega(\centi)} w(\filti,\centi) \mathbf{x}(\filti)
\end{equation}
where  the size of the search area $\Omega$ is determined by a parameter, and the patch size of $\mathbf{\hat{x}}$ and $\mathbf{x}$ are equal and given by $\pixpatch$. Since each pixel is part of multiple patches, the filtering procedure will estimate each pixel multiple times \cite{vitale2019gnlm}.

Note that the optical data does not enter directly into Eq.\ \eqref{eq:gnlm-pix}, which means that only SAR domain pixels are used for determining the filtered SAR image \cite{vitale2019gnlm}. It is only used to help determine the weights $w(\filti,\centi)$ in Eq.\ \eqref{eq:gnlm-pix}

The weight determining how much the filtering of a patch centred on pixel $\centi$ is influenced by patch centred on pixel $\filti$ can then be written as: %
\begin{equation}
\label{eq:weight-org}
    w(\filti,\centi) = C e^{ - \lambda \left[ \gamma \dissim_\text{SAR}(\filti,\centi) + (1-\gamma) \dissim_\text{OPT}(\filti,\centi) \right]}
\end{equation}
where $C$ is a normalising constant, $\dissim_\text{SAR}$ and $\dissim_\text{OPT}$ are patch-based dissimilarity measures in the SAR and optical domain respectively, $\lambda$ is an empirical weight parameter, and $\gamma \in \left[ 0, 1 \right]$ balances the emphasis on SAR versus optical dissimilarity.

For the optical domain, \cite{vitale2019gnlm} used the normalised sum of the squared Euclidean distance
\begin{equation}
\label{eq:opt-dissim}
    \dissim_\text{OPT}(\filti,\centi) = \frac{1}{\nband \npix} \sum_{\iband=1}^\nband \sum_{\ipix \in \pixpatch} \left[ \optpatch_\iband(\filti + \ipix) - \optpatch_\iband(\centi + \ipix) \right]^2
\end{equation}
where $\nband$ is the number of bands in the optical guide and $\npix$ is the number pixels in each patch determined by the set of spatial offsets $\pixpatch$.

The SAR dissimilarity measure used in \cite{vitale2019gnlm} was for multiplicative noise in single polarisation intensity data. To extend GNLM to PolSAR data, we chose to use a dissimilarity measure that utilised the polarimetric information. Since each pixel in the input SAR image is a complex scattering vector as defined in Eq.\ \eqref{eq:s-vec}, a dissimilarity measure between two such vectors can be defined as
\begin{equation}
  \label{eq:svec-dissim}
  \dissim_{( \mathbf{s}_\filti, \mathbf{s}_\centi)} = \frac{(\mathbf{s}_\centi - \mathbf{s}_\filti)^H (\mathbf{s}_\centi - \mathbf{s}_\filti) }{\mathbf{s}_\centi^H \mathbf{s}_\centi}
\end{equation}
If we sum this expression we can get a patch-based dissimilarity between the patches centred on pixel position $\centi$ and pixel position $\filti$
\begin{equation}
  \label{eq:sar-dissim}
  \dissim_\text{SAR}(\filti,\centi) = \frac{1}{\npix} \sum_{\ipix \in \pixpatch} \frac{(\mathbf{s}_{\centi +\ipix} - \mathbf{s}_{\filti +\ipix})^H (\mathbf{s}_{\centi +\ipix} - \mathbf{s}_{\filti +\ipix}) }{\mathbf{s}_{\centi +\ipix}^H \mathbf{s}_{\centi +\ipix}}
\end{equation}
where $\npix$ and $\pixpatch$ are defined as before.

By modifying Eq.\ \eqref{eq:gnlm-pix}, we can then find the polarimetric guided nonlocal means (PGNLM) estimate for the covariance matrix as:
\begin{equation}
\label{eq:pgnlm-covest}
  \mathbf{C}(\centi) = \sum\limits_{\filti \in \Omega(\centi)} w(\filti,\centi) \mathbf{s}_\filti \mathbf{s}_\filti^H
\end{equation}
Where the weight $w(\filti,\centi)$ is defined in Eq.\ \eqref{eq:weight-org}, $\dissim_\text{OPT}$ is given in Eq.\ \eqref{eq:opt-dissim}, and $\dissim_\text{SAR}$ in Eq.\ \eqref{eq:sar-dissim}.

Note that while each pixel intensity is estimated multiple times as it is a part of multiple patches in the case of single-channel intensity filtering in Eq.\ \eqref{eq:gnlm-pix}, the covariance matrix for pixel position $\centi$ is only estimated once. The estimate in Eq.\ \eqref{eq:pgnlm-covest} is based on pixels where the patch $\pixpatch$ centred on that pixel is sufficiently similar to the patch centred on the pixel position to be estimated, and weighted according to the dissimilarities in the SAR and optical domain.

\section{Results}
\label{sec:results}
For separating pixels with live crown foliage from those with defoliated crown, we train a random forest classifier with 200 trees on the filtered covariance matrices. For comparison we obtained the filtered covariance matrices using the boxcar, enhanced Lee, and intensity-driven adaptive-neighbourhood (IDAN) filters in SNAP. Both boxcar and enhanced Lee filters used a $5 \times 5$ window, while the adaptive-neighbourhood size for IDAN was 50.

The PGNLM parameters were set in a heuristic manner, following recommendations in \cite{vitale2019gnlm}. The search area was set to $39 \times 39$ pixels, while the size of the patches to be compared were $9 \times 9$. The balancing factor between SAR and optical dissimilarities, $\gamma$ in Eq.\ \eqref{eq:weight-org} was set to $0.85$, while $\lambda$ was $0.5$. Also the measures to discard unreliable predictors used in \cite{vitale2019gnlm} was employed.

All the polarimetric information is contained in the elements of $\mathbf{C}$, and we can further simplify the processing by extracting the properties with relevant information: $C_{11}$, $C_{22}$, $C_{33}$, $|C_{13}|$, and $\angle C_{13}$. Here, $C_{11}$, $C_{22}$, $C_{33}$ are the intensities in the HH, HV, and VV channels, respectively, and $C_{13}=|C_{13}|e^{j\angle C_{13}}$ is the cross-correlation between the complex scattering coefficients in the co-polarised channels HH and VV.

\begin{figure}[htb]
\begin{minipage}[b]{1.0\linewidth}
  \centering
  \includegraphics[width=1.0\linewidth, keepaspectratio]{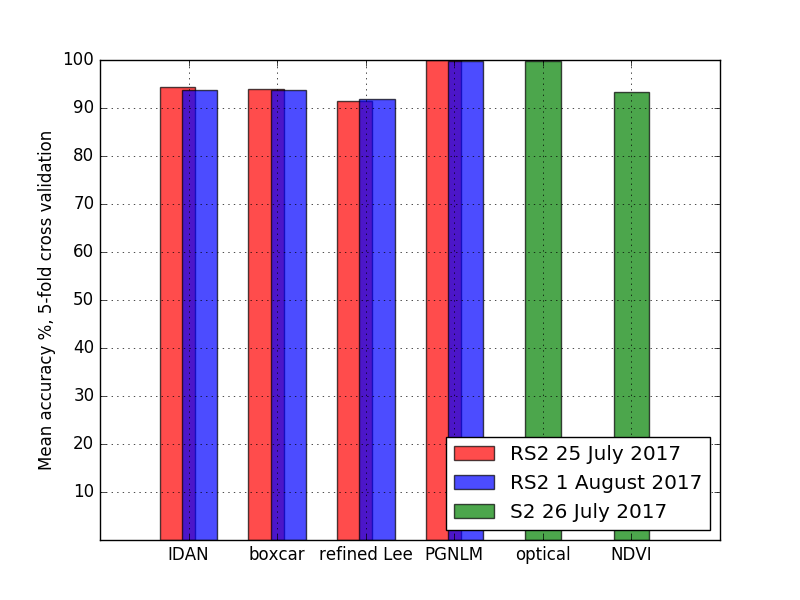}
\end{minipage}
\caption{Random forest classification accuracy.}
\label{fig:accbars}
\end{figure}

In addition, we compare with the classification result on the four-band optical Sentinel-2 subset used as the guide in PGNLM, as well as the NDVI. All data were divided into $\crossvalk$ parts for k-fold cross validation, and the average accuracy is reported. The result is seen in Figure \ref{fig:accbars}.

We see that PGNLM achieves 100\% accuracy for 25 July (red bar), 5.7 percentage points better than second best (IDAN). For the 1 August dataset (blue bar) PGNLM achieves 99.7\% accuracy, 6.0 percentage points better than second best (IDAN). The optical data (Sentinel-2 25 July), shown in green bars, achieves 99.9\% accuracy. As expected, the NDVI result is significantly lower, as it is based on only two out of four bands in the optical image.

\begin{figure}[htb]
\begin{minipage}[b]{1.0\linewidth}
  \centering
  \includegraphics[width=1.0\linewidth, keepaspectratio]{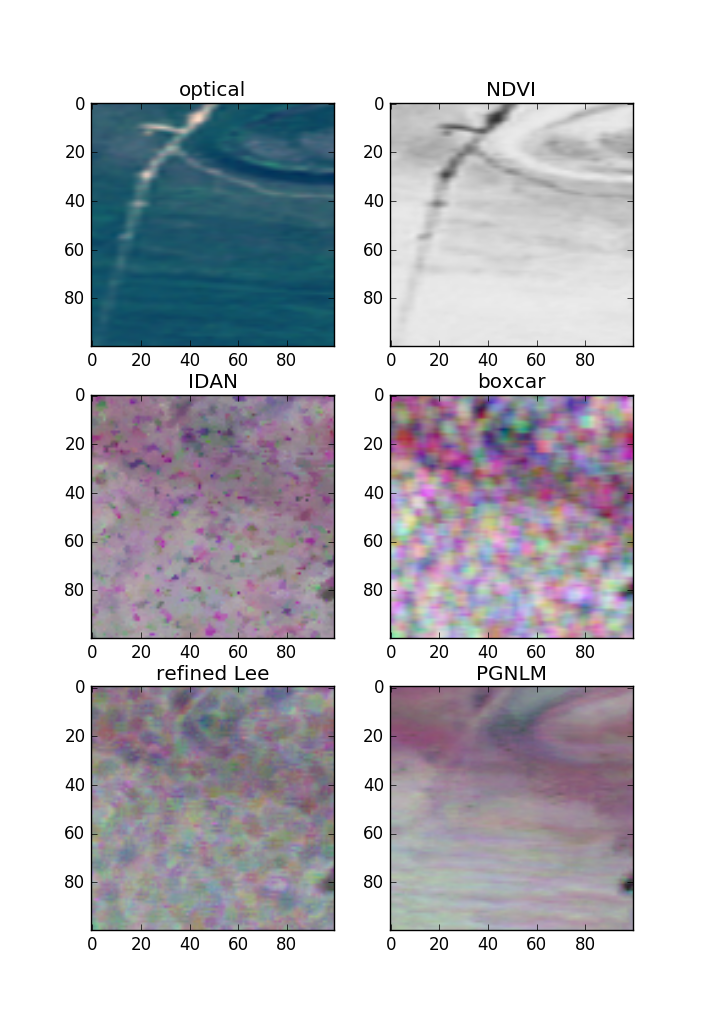}
\end{minipage}
\caption{Optical data (top row) and filtered SAR data.}
\label{fig:img-compare-6}
\end{figure}

Fig.\ \ref{fig:img-compare-6} shows a close-up of an area north of the easternmost live RA in Fig.\ \ref{fig:studysite}. For the SAR data, $C_{11}$, $C_{22}$, $C_{33}$ are normalised and shown in the red, green, and blue channels respectively.
The PGNLM algorithm achieves a significantly smoother result than the other filtering methods, and without any obvious filtering artefacts. This is not unexpected as it averages covariance matrix estimations from a large search area, while also discarding unreliable predictors.

\section{Conclusions and future work}
\label{sec:conclusion}
PGNLM filtered SAR data achieve the best accuracy results of the SAR filtering methods, and comparable to optical data.

The PGNLM algorithm contains quite a few parameters, that in various ways impact each other. Here they were set in a heuristic manner. For a better understanding, how to set the parameters for polarimetric guided nonlocal means should be explored, as was done for GNLM in \cite{vitale2019gnlm}. Also, applying PGNLM to standard datasets can help get a more accurate comparison of its performance relative to other polarimetric speckle filtering methods.

\bibliographystyle{IEEEtran}
{\small
\bibliography{bibliography}}

\end{document}